\documentclass[showpacs,nofootinbib,floatfix,superscriptaddress,aps]{revtex4}
\usepackage{graphicx}
\usepackage{epsfig}
\usepackage{bm}
\usepackage{amsfonts}
\usepackage[utf8]{inputenc}
\usepackage{amssymb}
\usepackage{float}
\usepackage{amsmath}
\usepackage{dcolumn}
\usepackage{latexsym}
\usepackage{multirow}
\usepackage{amsthm}
\usepackage{framed}
\usepackage{cancel}
\usepackage{mathtools}
\usepackage{tablefootnote}
\usepackage[colorlinks]{hyperref}
\usepackage[dvipsnames]{xcolor}
\hypersetup{
    breaklinks=true,
    pdfstartview={FitH},    
    colorlinks=true,       
    linkcolor=blue,          
    citecolor=red,        
    filecolor=magenta,      
    urlcolor=magenta,           
    anchorcolor=green,      
    linktocpage=true
}
\def\doi{http://doi.org}
\def\arxiv{http://arxiv.org/abs}
\makeatletter\def\Hy@Warning#1{}\makeatother
\newcommand{\labar}{\lower0.2ex\hbox{$\mathchar'26$}\mkern-10mu \lambda}

\begin{document}

\title{Extended Uncertainty Principle: A Deeper Insight into the Hubble Tension?}

\author{Kourosh Nozari}
\email[]{knozari@umz.ac.ir}
\affiliation{Department of Theoretical Physics, Faculty of Sciences, University of Mazandaran, 47416-95447 Babolsar, Iran}

\author{Sara Saghafi}
\email[]{s.saghafi@umz.ac.ir}
\affiliation{Department of Theoretical Physics, Faculty of Sciences, University of Mazandaran, 47416-95447 Babolsar, Iran}

\author{Milad Hajebrahimi}
\email[]{m.hajebrahimi@stu.umz.ac.ir}
\affiliation{Department of Theoretical Physics, Faculty of Sciences, University of Mazandaran, 47416-95447 Babolsar, Iran}

\begin{abstract}

The standard cosmological model, known as the $\Lambda$CDM model, has been successful in many respects, but it has some significant discrepancies, some of which have not been resolved yet. In measuring the Hubble-Lemaître parameter, there is an apparent discrepancy which is known as the Hubble tension, defined as differences in values of this parameter measured by the Type Ia Supernovae (SNeIa) data (a model-independent method) and by the Cosmic Microwave Background (CMB) radiation maps (a model-dependent method). Although many potential solutions have been proposed, the issue still remains unresolved. Recently, it was observed that the Hubble tension can be due to the concept of uncertainty in measuring cosmological parameters at large distance scales through applying the Heisenberg Uncertainty Principle (HUP) in cosmological setups. Extending this pioneering idea, in the present study we plan to incorporate the Extended Uncertainty Principle (EUP) containing a minimal fundamental measurable momentum (or equivalently, a maximal fundamental measurable length) as a candidate setup for describing large-scale effects of Quantum Gravity (QG) to address the Hubble tension and constrain the EUP length scale. In this regard, by finding a relevant formula for the effective photon rest mass in terms of the present-time value of the Hubble-Lemaître parameter, we see that discrepancies in the value of photon rest mass associated with the Hubble-Lemaître parameter values estimated from model-independent and model-dependent methods perhaps is the cause of Hubble tension. We show explicitly that the presence of a non-zero minimal uncertainty in momentum (or non-zero maximal uncertainty in position) measurements as a consequence of the large distance quantum gravity effect addresses the Hubble tension satisfactorily without having to invoke new physics. Moreover, we use the formula for the effective photon rest mass to find some relevant lower bounds on the EUP length scale.

\vspace{8 pt}

\end{abstract}


\fontsize{9}{10}
\selectfont
\maketitle

\section{Introduction}\label{sec:1}

In the vast array of cosmological scenarios, the $\Lambda$ Cold Dark Matter ($\Lambda$CDM) model \cite{Bull:2016stt,Perivolaropoulos:2022jda,DiValentino:2022fjm} has become the standard cosmological model due to its ability to accurately describe a wide range of astrophysical and cosmological phenomena. Despite $\Lambda$CDM successes, the model still contains significant unknowns, particularly in explaining fundamental aspects of the universe such as inflation \cite{Brout:1978ix,Guth:1981zm,Sato:1981qmu}, dark matter \cite{Rubin:1970zza,Trimble:1987ee}, and dark energy \cite{SupernovaSearchTeam:1998fmf,SupernovaCosmologyProject:1999vns}. These components, not predicted by theoretical first principles and unsupported by laboratory experiments, are solely based on cosmological and astrophysical observations. The $\Lambda$CDM model's simplistic assumptions about these ingredients liken it to a phenomenological effective theory, derived from theoretically motivated underlying frameworks that are yet to be fully understood. However, as observational precision improves, which causes physical cosmology to become a precision science, known as ``precision cosmology'' \cite{Bridle:2003yz,Primack:2004gb,Turner:2022gvw,Wu:2023dgy}, discrepancies within the model are becoming more apparent, suggesting the need for new physics beyond the $\Lambda$CDM model. In this regard, the most controversial discrepancy appears in the measurements of $H_{0}$ as the current (late-time) value of the Hubble-Lemaître parameter\footnote{The terminology of Hubble-Lemaître parameter is based on the International Astronomical Union (IAU) \cite{IAU:2018hll}, remembering the significant contributions made by Humason \cite{Humason:1929mlh,Humason:1931mlh,Hubble:1931eh}.}. This discrepancy is known as the Hubble tension (or $H_{0}$ tension) \cite{Bernal:2016gxb,Benetti:2018juy,Verde:2019ivm}.

To comprehend the concept of the Hubble tension, it is fruitful to know how $H_{0}$ value can be determined through two main methods: First, by measuring the luminosity distance and the recessional velocity resulting from the cosmic expansion of established standard candles, and then calculating the proportionality factor according to the Hubble-Lemaître law. This method is independent of any specific model; Second, by utilizing early universe probes, e.g., Cosmic Microwave Background (CMB) radiation maps, and assuming a specific model for the universe's expansion history, such as the $\Lambda$CDM model. Accordingly, the observations of the Planck Satellite Mission (PSM) from precise CMB radiation maps incorporated with Baryonic Acoustic Oscillations (BAO) data demonstrate that $H_{0}^{(\mathrm{PSM})}=67.4\pm 0.5\,\,\mathrm{km\cdot s^{-1}\cdot Mpc^{-1}}$ by assuming the standard $\Lambda$CDM model \cite{Planck:2018vyg}. Whereas the observations of the Hubble Space Telescope (HST) from the Type Ia Supernovae (SNeIa) show that $H_{0}^{(\mathrm{HST})}=74.03\pm 1.42\,\,\mathrm{km\cdot s^{-1}\cdot Mpc^{-1}}$ by calibrating the SNeIa with the Cepheids (also called Cepheid Variables) \cite{Riess:2019cxk}. These values show the Hubble tension at $4.4\sigma$ level \cite{Riess:2019cxk}. Also, more precise investigations of HST data recently show that $H_{0}^{(\mathrm{HST})}=73.04\pm 1.04\,\,\mathrm{km\cdot s^{-1}\cdot Mpc^{-1}}$, which causes the tension to become at $5\sigma$ level \cite{Riess:2022jrx}. A number of research studies have been conducted to address the Hubble tension \cite{DiValentino:2021izs,Schoneberg:2021qvd,Poulin:2023lkg,Dainotti:2023yrk,Clifton:2024mdy,
Vagnozzi:2022tjv,Vagnozzi:2021gjh,Vagnozzi:2020ezj,DiValentino:2020ffd,DiValentino:2020jae}. Nevertheless, the Hubble tension is still an open problem. Recently, the James Webb Space Telescope (JWST), deepened the mystery of the Hubble tension. The high-resolution observational data from the JWST confirm $H_{0}$ value measured by HST \cite{Riess:2023bfx}.

In 2020, Capozziello et al. \cite{Capozziello:2020nyq} proposed a new approach (later developed in Ref. \cite{Spallicci:2021kye}) to address the Hubble tension by associating it with the concept of measurement within a cosmological framework by incorporating the Heisenberg Uncertainty Principle (HUP), i.e., $\Delta x\Delta p\geq\hbar/2$ where $\hbar$ is the reduced Planck constant and $\Delta x$ and $\Delta p$ are the uncertainties in measuring position and momentum, respectively. Their idea was based on two major considerations. First, they assumed that luminosity distance deriving from a cosmographic Taylor series up to the second order is equivalent to the reduced Compton wavelength. Second, inspired by theories introducing massive photons \cite{Broglie:1936del,Stueckelberg:1957ecg,Boulware:1970zc,Guendelman:1979es,Goldhaber:2010xy,Bonetti:2017vrq,Bonetti:2018toa}, they assigned a rest mass to low-energy photons corresponding with the large Compton wavelength of the observed universe as a function of $H_{0}$. Accordingly, they measured the photon mass to address the Hubble tension at $4.4\sigma$ without demanding new physics. They found that the Hubble tension could be a result of uncertainty in measuring cosmological parameters in cosmic scales. However, the photon rest mass they computed was much smaller than what is predicted by experiments and observations, which shows the demand of going beyond the Heisenberg uncertainty limit in the framework of their idea to reconcile the Hubble tension.

The ultimate Quantum Gravity (QG) theory -- a framework for describing gravity in small-scale regimes where quantum features are considerable- which provides the possibility of investigating the transition of gravitational interaction from classical to quantum behavior, has not yet been formulated. However, black hole physics \cite{Casadio:2014pia} together with all phenomenological approaches to QG proposal including String Theory (ST) \cite{Amati:1987wq,Gross:1988ar,Amati:1989tn,Yoneya:1989ai,Konishi:1990wk}, Loop Quantum Gravity (LQG) \cite{Rovelli:1995ge,Ashtekar:1997eg,Ashtekar:1998fb,Modesto:2009jz}, Asymptotically Safe Quantum Gravity (ASQG) \cite{Lauscher:2005qz,Reuter:2006bb,Percacci:2010af,Eichhorn:2019yfc}, Noncommutative Geometry (NG) \cite{Seiberg:1999vs,Connes:2000by}, and gedankenexperiments (thought experiments) proposed to unify Quantum Mechanics (QM) and General Relativity (GR) \cite{Hossenfelder:2012jw} predict the existence of a minimal fundamental measurable length of order of the Planck length, $l_{\mathrm{pl}}$. To incorporate this minimal length in ordinary quantum mechanics, one has to modify the HUP to result in the so-called Generalized (Gravitational) Uncertainty Principle (GUP). A GUP setup includes the gravitationally induced position uncertainty associated with the minimal fundamental length. On the other hand, the Doubly (Deformed) Special Relativity (DSR) \cite{Amelino-Camelia:2002stu,Cortes:2005qn} has developed GUP setups to consider both the minimal measurable length and its energy-scale counterpart, i.e., a maximal fundamental measurable momentum (energy). Many works have been devoted to studying the phenomenological features of various GUP setups and their applications (see, e.g., Refs. \cite{Kempf:1994su,Pedram:2011xj,Nozari:2012gd,Nozari:2012nf,Nozari:2019nxj,Bosso:2023aht}). However, the quantum mechanical uncertainty relations are also affected by the curvature of spacetime geometry. The absence of a notion of a plane wave on a generally curved spacetime has been generally believed to occur in large-scale physics, where the curvature of spacetime becomes significant \cite{Hinrichsen:1996mf,Kempf:1997ss}. This fact implies the existence of a limit to the precision of momentum measurements, i.e., there is a minimal fundamental measurable momentum. In this regard, the large-scale induced correction on the HUP leads to the framework of the Extended Uncertainty Principle (EUP) containing minimal fundamental measurable momentum. Unlike the GUP, the studies on the EUP frameworks are limited. Some recent studies have focused on the phenomenology of the EUP, including Refs. \cite{Illuminati:2021wfq,Okcu:2022sio,Lu:2019wfi,Mureika:2019gxl,Dabrowski:2019wjk,Gine:2020izd,Oubagha:2023ghx,Okcu:2022iwl,Chung:2019oqo,Schurmann:2019jxe}. Interestingly, the bounds from the Hubble tension on the GUP and EUP modification parameters are found in Ref. \cite{Aghababaei:2021gxe}. Furthermore, in a recent study \cite{Trivedi:2022bis}, Trivedi found that including GUP effects and extra dimensions into the Capozziello et al. proposal separately, as two new physical considerations, does not resolve the Hubble tension. The deduced photon rest mass was determined to be an irrelevant mass scale. Therefore, Trivedi concludes that only the HUP itself can effectively address the Hubble tension within the framework of the original Capozziello et al. proposal \cite{Capozziello:2020nyq}.

As previously mentioned, the Hubble tension based on the Capozziello et al. proposal \cite{Capozziello:2020nyq} can be reconciled beyond the Heisenberg indeterminacy limit. In this study, we aim to examine if the Capozziello et al. scheme affected by the EUP effects as a new physical consideration, can elucidate the Hubble tension. We first find the reduced Compton wavelength affected by a general EUP setup, which makes the reduced Compton wavelength shorter compared to its ordinary version or even when it is influenced by the GUP \cite{Trivedi:2022bis}. This can provide a situation in which it is possible to go beyond the Heisenberg indeterminacy limit. Then, through a cosmographic Taylor series up to the second order, we set the luminosity distance equivalent to the EUP-modified reduced Compton wavelength to measure the photon rest mass as a function of $H_{0}$. The notion of equivalency of the luminosity distance with the reduced Compton wavelength is not a straightforward one and may be deemed ad hoc. Although it can be argued that all cosmological distances are linked by powers of $(1+z)$ where $z$ is redshift, this relationship is not generic as it depends on the assumption of the Etherington distance-duality connection \cite{Etherington:1933led}. The purpose of considering this equivalence is to examine whether an analysis grounded in similar principles remains valid when incorporating fundamental quantum gravitationally motivated modifications, specifically the effects of EUP. The paper is structured as follows. In Section \ref{sec:2}, we provide details on the theoretical framework, including the EUP setup, the EUP-corrected reduced Compton wavelength, the luminosity distance formula, and a brief review on the Capozziello et al. proposal \cite{Capozziello:2020nyq}. In Section \ref{sec:3}, following the Capozziello et al. proposal \cite{Capozziello:2020nyq}, we consider the equivalency between the EUP-corrected reduced Compton wavelength and the luminosity distance up to second order to examine the possibility of addressing the Hubble tension in the setup and to find some bounds on the EUP modification length scale. Finally, in Section \ref{sec:4}, we end with conclusions.

\section{Theoretical Framework}\label{sec:2}

This section commences with an overview on the setup of the EUP. Subsequently, we proceed to reformulate the Compton wavelength within the EUP framework. We also present a description on the cosmological luminosity distance formula. Finally, we review the main idea of the Capozziello et al. proposal \cite{Capozziello:2020nyq}.

\subsection{EUP formulation}

The simplest deformed commutation relation between position operator $\hat{x}$ (with the expectation value $x$) and momentum operator $\hat{p}$ for the EUP setup can be determined as follows
\begin{equation}\label{cmre}
[\hat{x},\hat{p}]=\mathrm{i}\hbar\left(1+\frac{\beta_{0}}{L^{2}}x^{2}\right)\,,
\end{equation}
where the parameter $L$ represents a fundamental length scale and $\beta_{0}$ is a dimensionless quantity associated with the EUP setup, which is considered to be of the order of unity. One may define $\beta=\frac{\beta_{0}}{L^{2}}$ as the EUP parameter utilized in Sec. \ref{subsecc} as a major parameter. Therefore, the simplest EUP formula can be deduced from the EUP-deformed commutation relation \eqref{cmre} as follows \cite{Bambi:2008ty}
\begin{equation}\label{eupme}
\Delta{x}\Delta{p}\geq\hbar\left(1+\frac{\beta_{0}}{L^{2}}\left(\Delta{x}\right)^{2}\right)\,.
\end{equation}
The EUP formula \eqref{eupme} indicates the presence of a non-zero minimal uncertainty in momentum measurements within the system as follows
\begin{equation}\label{nzmiunmo}
\left(\Delta{p}\right)_{\mathrm{min}}=\frac{2\hbar\sqrt{\beta_{0}}}{L}\,,
\end{equation}
which is referred to as the natural IR cutoff. In simpler terms, Eq. \eqref{nzmiunmo} represents the minimum measurable momentum, which in turn corresponds to the largest possible length scale.

The fundamental length $L$ is the key characteristic of the EUP framework. Several works have been conducted to constrain $L$ through astrophysical tests; some of which are listed in Table \ref{Table1}. On the other hand, one can set the EUP fundamental length $L$ as the Hubble length so that $L=L_{\mathrm{H}}=\frac{c}{H_{0}}$ where $c$ is the speed of light in vacuum \cite{Mureika:2019gxl}. It is worth mentioning that the EUP fundamental length can also be attributed to the radius of Anti-de Sitter (AdS) spacetime so that $L_{\mathrm{AdS}}^{2}=\frac{3}{\Lambda}$ where $\Lambda$ is the cosmological constant \cite{Hassanabadi:2020osz}. However, this cannot be considered in the present study since the Hubble tension is associated with the Friedmann-Lemaître-Robertson-Walker (FLRW) universe.
\begin{table}[htb]
  \centering
  \caption{\label{Table1}\emph{Lower bounds on the fundamental length $L$.}}
  \def\arraystretch{1.1}
  \begin{tabular}{l|l|l}
  \hline\hline
  Test & Bound on $L\,\mathrm{(m)}$ & Ref. \\
  \hline
  Lunar laser ranging & $>4.9\times 10^{5}$ & \cite{Illuminati:2021wfq} \\
  Perihelion shift of Mercury’s orbit & $>5.8\times 10^{5}$ & \cite{Okcu:2022sio} \\
  Shapiro time delay & $>9\times 10^{5}$ & \cite{Okcu:2022sio} \\
  Light deflection & $>9.1\times 10^{5}$ & \cite{Lu:2019wfi} \\
  Torsion pendulum & $>1.5\times 10^{7}$ & \cite{Illuminati:2021wfq} \\
  Perihelion precession & $>5.05\times 10^{9}$ & \cite{Illuminati:2021wfq} \\
  Strong lensing (Sgr A*\footnotemark) & $>2\times 10^{10}$ & \cite{Lu:2019wfi} \\
  Precession of S2 star's orbit & $>4\times 10^{10}$ & \cite{Okcu:2022sio} \\
  Binary pulsar & $>1.4\times 10^{11}$ & \cite{Illuminati:2021wfq} \\
  Solar-spin precession & $>5.05\times 10^{11}$ & \cite{Illuminati:2021wfq} \\
  EUP black holes & $>10^{12}$ & \cite{Mureika:2019gxl} \\
  Strong lensing (M87*\footnotemark) & $>3\times 10^{13}$ & \cite{Lu:2019wfi} \\
  \hline\hline
  \end{tabular}
  \footnotetext[1]{\,The abbreviation of Sagittarius A* that is the supermassive black hole at the center of the Milky Way.}
  \footnotetext[2]{\,The abbreviation of Messier 87* that is the supermassive black hole at the center of the Messier 87 (M87) galaxy.}
\end{table}

\subsection{EUP-modified reduced Compton wavelength}

The Compton wavelength is a distinctive quantum mechanical attribute of a particle that is precisely defined as the wavelength of a photon with an energy equivalent to the rest energy of the particle. The Compton wavelength represents an essential constraint on the ability to precisely determine the position of a particle, considering the principles of special relativity and quantum mechanics. Furthermore, this observation establishes that the Compton wavelength serves as a distance cutoff below which the effects of quantum field theory become significant. The reduced Compton wavelength is described as follows
\begin{equation}\label{rcwl}
\labar_{\mathrm{C}}=\frac{\hbar}{m_{0}c}\,,
\end{equation}
where $m_{0}$ is the rest mass of the particle. Now, one can suppose the reduced Compton wavelength as the uncertainty in position measurement, i.e., $\labar_{\mathrm{C}}=\Delta x$. Similarly, one can assume $m_{0}c=\Delta p$ as the uncertainty in momentum measurement \cite{Capozziello:2020nyq}. Accordingly, isolating $\Delta x$ in Eq. \eqref{eupme} and implementing the assumptions mentioned above result in the EUP-modified reduced Compton wavelength as follows
\begin{equation}\label{eupmcwl}
\labar_{_{\mathrm{EUP}}}=\frac{m_{\gamma}cL^{2}}{2\hbar\beta_{0}}\left(1-\sqrt{1-4\beta_{0}\left(\frac{l_{\mathrm{pl}}\,m_{\mathrm{pl}}}{m_{\gamma}L}\right)^{2}}\right)\,,
\end{equation}
where $m_{\mathrm{pl}}$ is the Planck mass and also for the aims of the present study as Ref. \cite{Capozziello:2020nyq}, we set $m_{0}\equiv m_{\gamma}$ in which $m_{\gamma}$ is the rest mass of photon. In the limit $\beta_{0}\rightarrow0$, one can recover the ordinary reduced Compton wavelength as described in Eq. \eqref{rcwl}.

\subsection{Luminosity distance}

Here, we proceed to consider the cosmographic or cosmokinematic approach \cite{Cattoen:2007id,Guimaraes:2011mw} which can be defined as a framework of physical cosmology in which minimal assumptions related to dynamical aspects of the universe are taken into account. In this sense, one can express the standard relation for luminosity distance in terms of redshift truncated at the second order through a simple Taylor series expansion as follows \cite{Cattoen:2007sk,Cattoen:2008th}
\begin{equation}\label{ludisec}
d_{\mathrm{L}}(z)=\frac{c}{H_{0}}\left(z+\frac{1}{2}(1-q_{0})z^{2}\right)+\mathcal{O}(z^{3})\,,
\end{equation}
where the present day value of the deceleration parameter is
\begin{equation}\label{cosgrse}
q_{0}=-\left(\frac{1}{a(t_{0})H_{0}^{2}}\right)\frac{\mathrm{d}^{2}a(t)}{\mathrm{d}t^{2}}\bigg|_{t=t_{0}}\,,
\end{equation}
in which $a(t)$ is the scale factor of the universe and $t_{0}$ is the current cosmic time. It is worth mentioning that according to the Taylor series expansion, the luminosity distance in terms of redshift described in Eq. \eqref{ludisec} is valid just for small redshift values ($z\lesssim 1$) due to some convergence issues at high redshift as can be seen in Eq. \eqref{ludisec} \cite{Cattoen:2007sk,Cattoen:2008th}.

Based on the values of $q_{0}$, three situations can be occurred \cite{Aviles:2014rma}. First, the situation $q_{0}>0$ signifies an expanding universe that is presently in a decelerating phase. This occurs in a universe dominated by matter or any pressureless barotropic fluid. Nevertheless, recent observations contradict $q_{0}>0$ for the current epoch of cosmic evolution. This situation was significant in early universe when matter was dominant over other ingredients of the universe. Second, the situation $-1<q_{0}<0$ describes a universe, which is currently undergoing accelerated expansion, which is the case for the universe we now live in. It is believed that the universe is primarily influenced by a type of antigravitational substance known as dark energy. Furthermore, cosmography approach validates these features without proposing a specific model for the evolution of dark energy. Third, the situation $q_{0}=-1$ represents that the primary source of energy in the cosmos is controlled by a cosmic element with a consistent energy density known as de Sitter fluid that remains unchanged as the universe expands. This scenario was characteristic of the inflationary period in the very early universe. There are a number of constraints on the values of $q_{0}$. For instance, in Ref. \cite{Capozziello:2018jya}, the value $q_{0}=-0.644\pm 0.223$ is reported utilizing joint Pantheon data for SNeIa, BAO, and time-delay measurements by H0LiCOW, together with angular diameter distances measured by implementing water megamasers. As explained above, this value is in agreement with the late-time positively accelerated expansion phase of the universe.

\subsection{Equivalence between Compton wavelength and luminosity distance}

In the initial version of the proposal by Capozziello et al. \cite{Capozziello:2020nyq}, the reduced Compton wavelength of the observable universe was equated to the luminosity distance in terms of redshift truncated at the second order described in Eq. \eqref{ludisec}, while taking into consideration the condition $z\sim 1$. As a result, this configuration led to determining the effective rest mass of photon as a function of $H_{0}$ to address the Hubble tension. However, in the original proposal by Capozziello et al. \cite{Capozziello:2020nyq}, when both $H_{0}^{(\mathrm{PSM})}$ and $H_{0}^{(\mathrm{HST})}$ are taken into account, separately, the effective rest mass of photon was found to the order of $m_{\mathrm{\gamma}}\sim 10^{-69}\,\mathrm{kg}$, which is significantly lower than the most precise upper limits on the photon mass that have been recently reported. The Particle Data Group (PDG) \cite{ParticleDataGroup:2022pth} employing data from the solar system \cite{Retino:2016gga} reported $m_{\mathrm{\gamma}}<10^{-54}\,\mathrm{kg}$ as the upper bound on the rest mass of photons, while experimental tests \cite{Williams:1971ms} set the bound as $m_{\mathrm{\gamma}}<1.6\times 10^{-50}\,\mathrm{kg}$. On the other hand, Capozziello et al. \cite{Capozziello:2020nyq} computed the actual change of the photon rest mass as
\begin{equation}\label{agmg}
\Delta m_{\gamma}=m_{\gamma}^{(\mathrm{HST})}-m_{\gamma}^{(\mathrm{PSM})}\,,
\end{equation}
and the actual change of the Hubble-Lemaître parameter as
\begin{equation}\label{aghp}
\Delta H_{0}=H_{0}^{(\mathrm{HST})}-H_{0}^{(\mathrm{PSM})}\,.
\end{equation}
Then, they found the classical relative change of the photon rest mass to the form of $\frac{\Delta
m_{\gamma}}{m_{\gamma}}$ and the classical relative change of the Hubble-Lemaître parameter such that $\frac{\Delta H_{0}}{H_{0}}$. Capozziello et al. \cite{Capozziello:2020nyq} discovered that $\frac{\Delta m_{\gamma}}{m_{\gamma}}=\frac{\Delta H_{0}}{H_{0}}\approx 0.1$. Consequently, these observations demonstrate that the original proposal by Capozziello et al. \cite{Capozziello:2020nyq} establishes that although the discrepancy in the Hubble-Lemaître parameter can potentially be resolved by measuring the photon mass beyond the HUP limit, this discrepancy may actually stem from uncertainties in measurements at cosmological scales without having to invoke new physics. 

It is important to highlight that around $z\sim 1$, the kinematic interpretation of the Hubble-Lemaître law may be suppressed by the dynamical explanation derived from the Friedmann equations. Consequently, the cosmographic Taylor series expansion used to relate the luminosity distance with redshift needs to be limited to the second order, as shown in Eq. \eqref{ludisec}. This adjustment is necessary because the kinematic and dynamic aspects of the Hubble-Lemaître law come into conflict at the second order truncation in the cosmographic Taylor series expansion. The assumption of $z\sim 1$ is made in the original version of the Capozziello et al. proposal \cite{Capozziello:2020nyq}.

In a recent study \cite{Trivedi:2022bis}, Trivedi has expanded upon the original proposal by Capozziello et al. \cite{Capozziello:2020nyq} by incorporating both GUP and compactified higher dimensions, separately. These enhanced versions of the Capozziello et al. proposal indicated that modifications based on GUP and the inclusion of compactified higher dimensions do not address the Hubble tension through the concept of measurement. A similar conclusion is drawn in \cite{Vagnozzi:2023nrq}, demonstrating that resolving the Hubble tension necessitates more than just early-time new physics, like GUP effects, alone.

\section{Results}\label{sec:3}

In this section, following the proposal by Capozziello et al. \cite{Capozziello:2020nyq}, we aim to find the effective rest mass of photon in the EUP setup. Thus, we take into account the luminosity distance \eqref{ludisec} as the EUP-modified reduced Compton wavelength \eqref{eupmcwl}, such that
\begin{equation}\label{ludcow}
d_{\mathrm{L}}(z)=\labar_{_{\mathrm{EUP}}}\Rightarrow\frac{c}{H_{0}}\left(z+\frac{1}{2}(1-q_{0})z^{2}\right)=
\frac{m_{\gamma}cL^{2}}{2\hbar\beta_{0}}\left(1-\sqrt{1-4\beta_{0}\left(\frac{l_{\mathrm{pl}}\,m_{\mathrm{pl}}}{m_{\gamma}L}\right)^{2}}\right)\,,
\end{equation}
to find the effective rest mass of photon in the EUP setup as follows
\begin{equation}\label{ermp}
m_{\gamma}=\frac{H_{0}(l_{\mathrm{pl}}\,m_{\mathrm{pl}})^{2}}{\hbar}\left(z+\frac{1}{2}(1-q_{0})z^{2}\right)^{-1}
+\frac{\beta_{0}\hbar}{H_{0}L^{2}}\left(z+\frac{1}{2}(1-q_{0})z^{2}\right)\,.
\end{equation}
As seen from Eq. \eqref{ermp}, the effective rest mass of photon in the EUP setup depends on multiple parameters including $H_{0}$, $q_{0}$, $z$, $\beta_{0}$, and specifically $L$.

On the other hand, using Eqs. \eqref{agmg} and \eqref{aghp}, we will implement the absolute change of the photon rest mass $|\Delta m_{\gamma}|$ and the absolute change of the Hubble-Lemaître parameter $|\Delta H_{0}|$ to find the arithmetic mean change of the photon rest mass
\begin{equation}\label{amcmg}
\frac{|\Delta m_{\gamma}|}{\bar{m}_{\gamma}}\,,
\end{equation}
and the arithmetic mean change of the Hubble-Lemaître parameter
\begin{equation}\label{amchp}
\frac{|\Delta H_{0}|}{\bar{H}_{0}}\,,
\end{equation}
where we have defined
\begin{equation}\label{mmghlp}
\bar{m}_{\gamma}=\frac{1}{2}\left(m_{\gamma}^{(\mathrm{HST})}+m_{\gamma}^{(\mathrm{PSM})}\right)\,,\qquad
\bar{H}_{0}=\frac{1}{2}\left(H_{0}^{(\mathrm{HST})}+H_{0}^{(\mathrm{PSM})}\right)\,.
\end{equation}
We will use the arithmetic mean change of the photon rest mass and the Hubble-Lemaître parameter to check whether the Hubble tension is addressed.

Utilizing the effective rest mass of photon and the arithmetic mean change formulae, we are planning to focus on three different purposes in this EUP framework. Firstly, we want to find the mass of photon associated with different values of $L$ by considering both $H_{0}$ values, separately. Precisely speaking, our first aim is to verify whether the EUP setup addresses the Hubble tension by considering previously reported values of $L$ through finding the associated photon mass value. As the second purpose, we set $L_{\mathrm{H}}=\frac{c}{H_{0}}$ as the Hubble length to see whether the EUP setup addresses the Hubble tension. Finally, as the third purpose, we aim to constrain $\beta$ for both $H_{0}$ values, separately using the previously reported values for photon mass. This can demonstrate that for what accurate value of $L$ associated with both $H_{0}$ values separately, the Hubble tension may be addressed in the EUP framework. In rest of the paper, as mentioned previously, we set $z=1$ and $q_{0}=-0.644$ \cite{Capozziello:2018jya}.

\subsection{Various values of $L$ to address Hubble tension}

Here, we aim to implement the previously reported values of $L$, which are listed in Table \ref{Table1}, to evaluate the effective rest mass of photon \eqref{ermp} to address the Hubble tension utilizing the arithmetic mean change formula. To achieve this, we first insert the numerical values of the constant parameters in Eq. \eqref{ermp} with $\beta_{0}=1$ to find the following result
\begin{equation}\label{ermpnu}
m_{\gamma}=6.433\times 10^{-52}H_{0}+\frac{1.922\times 10^{-34}}{H_{0}L^{2}}\,,
\end{equation}
which is in terms of $H_{0}$ and $L$.

For three values of $L$ mentioned in Table \ref{Table1}, we evaluate Eqs. \eqref{amcmg}-\eqref{ermpnu} gathered from implementing $H_{0}^{(\mathrm{PSM})}=67.4\,\,\mathrm{km\cdot s^{-1}\cdot Mpc^{-1}}$ \cite{Planck:2018vyg} with $H_{0}^{(\mathrm{HST})}=74.03\,\,\mathrm{km\cdot s^{-1}\cdot Mpc^{-1}}$ \cite{Riess:2019cxk} demonstrating the Hubble tension of $4.4\sigma$ level and also with $H_{0}^{(\mathrm{HST})}=73.04\,\,\mathrm{km\cdot s^{-1}\cdot Mpc^{-1}}$ \cite{Riess:2022jrx} describing the tension at $5\sigma$ level. It should be noted that we dropped the errors in these values for the sake of simplicity. Finally, the outcomes are shown in Table \ref{Table2}. As seen from Table \ref{Table2}, none of the $L$ values up to $\sim 10^{13}\,\mathrm{km}$ can lead to the reported upper bounds on the rest mass of photon, i.e., of the order of $\sim 10^{-54}\,\mathrm{kg}$ \cite{ParticleDataGroup:2022pth,Retino:2016gga} or $\sim 10^{-50}\,\mathrm{kg}$ \cite{Williams:1971ms}. Moreover, despite the results of the proposal by Capozziello et al. \cite{Capozziello:2020nyq}, the resulted rest mass of photon associated with $H_{0}^{(\mathrm{PSM})}$ is greater than the corresponding result for $H_{0}^{(\mathrm{HST})}$ at $4.4\sigma$ or $5\sigma$ level. Furthermore, we see from Table \ref{Table2} that increasing the value of $L$ results in tending the achieved rest mass of photon to the aforementioned reported upper bounds on it. This demonstrates that with greater values of $L$, the reported rest mass of photon of the order of $\sim 10^{-54}\,\mathrm{kg}$ \cite{ParticleDataGroup:2022pth,Retino:2016gga} or $\sim 10^{-50}\,\mathrm{kg}$ \cite{Williams:1971ms} apparently can be resulted. We aim to find this value of $L$, subsequently. On the other hand, we reported the values of the arithmetic mean changes of the photon rest mass and the Hubble-Lemaître parameter in Table \ref{Table2} for both $H_{0}^{(\mathrm{PSM})}=67.4\,\,\mathrm{km\cdot s^{-1}\cdot Mpc^{-1}}$ \cite{Planck:2018vyg} with $H_{0}^{(\mathrm{HST})}=74.03\,\,\mathrm{km\cdot s^{-1}\cdot Mpc^{-1}}$ \cite{Riess:2019cxk} depicting the Hubble tension at $4.4\sigma$ level and with $H_{0}^{(\mathrm{HST})}=73.04\,\,\mathrm{km\cdot s^{-1}\cdot Mpc^{-1}}$ \cite{Riess:2022jrx} explaining the tension at $5\sigma$ level, respectively. We discover that the values of the arithmetic mean changes of the photon rest mass and the Hubble-Lemaître parameter are approximately equal for each value of $L$ corresponding with the tension at both $4.4\sigma$ and $5\sigma$. This means that the Hubble tension apparently is an outcome of EUP constraint on measuring $H_{0}$. Therefore, despite the previously reported values of $L$ that cannot lead to appropriate upper bounds on the effective photon rest mass, our setup provides an explanation for the Hubble tension. This explanation is based on the QG effects in the late-time epochs within the framework of EUP. The presence of non-zero minimal uncertainty in momentum (or non-zero maximal uncertainty in position) measurements can account for this explanation without having to invoke new physics.
\begin{table}[htb]
  \centering
  \caption{\label{Table2}\emph{The resulted photon rest mass and the arithmetic mean change associated with various values of $L$ in the EUP setup.}}
  \def\arraystretch{1.55}
  \begin{tabular}{l|l||l||l|l|l}
  \hline\hline
  Test and Ref.            & $L\,\mathrm{(m)}$                & $H_{0}\,\mathrm{(km\cdot s^{-1}\cdot Mpc^{-1})}$                                  & $m_{\gamma}\,\mathrm{(kg)}$ & $\frac{|\Delta m_{\gamma}|}{\bar{m}_{\gamma}}$ & $\frac{|\Delta H_{0}|}{\bar{H}_{0}}$ \\
  \hline
  \multirow{3}{*}{Torsion pendulum \cite{Illuminati:2021wfq}} & \multirow{3}{*}{$1.5\times 10^{7}$} & $H_{0}^{(\mathrm{PSM})}=67.4$ \cite{Planck:2018vyg}              & $m_{\gamma}^{(\mathrm{PSM})}\approx 3.912\times 10^{-31}$ & & \\
                           &                                                                        & $H_{0}^{(\mathrm{HST})}=74.03\,(\mathrm{at}\,4.4\sigma)$ \cite{Riess:2019cxk} & $m_{\gamma}^{(\mathrm{HST})}\approx 3.561\times 10^{-31}$ & $\approx 0.09$ & $\approx 0.09$ \\
                           &                                                                        & $H_{0}^{(\mathrm{HST})}=73.04\,(\mathrm{at}\,5\sigma)$ \cite{Riess:2022jrx}  & $m_{\gamma}^{(\mathrm{HST})}\approx 3.609\times 10^{-31}$ & $\approx 0.08$ & $\approx 0.08$ \\
  \hline
  \multirow{3}{*}{Strong lensing (Sgr A*) \cite{Lu:2019wfi}}  & \multirow{3}{*}{$2\times 10^{10}$}  & $H_{0}^{(\mathrm{PSM})}=67.4$ \cite{Planck:2018vyg}              & $m_{\gamma}^{(\mathrm{PSM})}\approx 2.200\times 10^{-37}$ & & \\
                           &                                                                        & $H_{0}^{(\mathrm{HST})}=74.03\,(\mathrm{at}\,4.4\sigma)$ \cite{Riess:2019cxk} & $m_{\gamma}^{(\mathrm{HST})}\approx 2.003\times 10^{-37}$ & $\approx 0.09$ & $\approx 0.09$ \\
                           &                                                                        & $H_{0}^{(\mathrm{HST})}=73.04\,(\mathrm{at}\,5\sigma)$ \cite{Riess:2022jrx}  & $m_{\gamma}^{(\mathrm{HST})}\approx 2.030\times 10^{-37}$ & $\approx 0.08$ & $\approx 0.08$ \\
  \hline
  \multirow{3}{*}{Strong lensing (M87*) \cite{Lu:2019wfi}}    & \multirow{3}{*}{$3\times 10^{13}$}  & $H_{0}^{(\mathrm{PSM})}=67.4$ \cite{Planck:2018vyg}              & $m_{\gamma}^{(\mathrm{PSM})}\approx 9.779\times 10^{-44}$ & & \\
                           &                                                                        & $H_{0}^{(\mathrm{HST})}=74.03\,(\mathrm{at}\,4.4\sigma)$ \cite{Riess:2019cxk} & $m_{\gamma}^{(\mathrm{HST})}\approx 8.903\times 10^{-44}$ & $\approx 0.09$ & $\approx 0.09$ \\
                           &                                                                        & $H_{0}^{(\mathrm{HST})}=73.04\,(\mathrm{at}\,5\sigma)$ \cite{Riess:2022jrx}  & $m_{\gamma}^{(\mathrm{HST})}\approx 9.023\times 10^{-44}$ & $\approx 0.08$ & $\approx 0.08$ \\
  \hline\hline
  \end{tabular}
\end{table}

\begin{figure}[htb]
\centering
\includegraphics[width=0.7\textwidth]{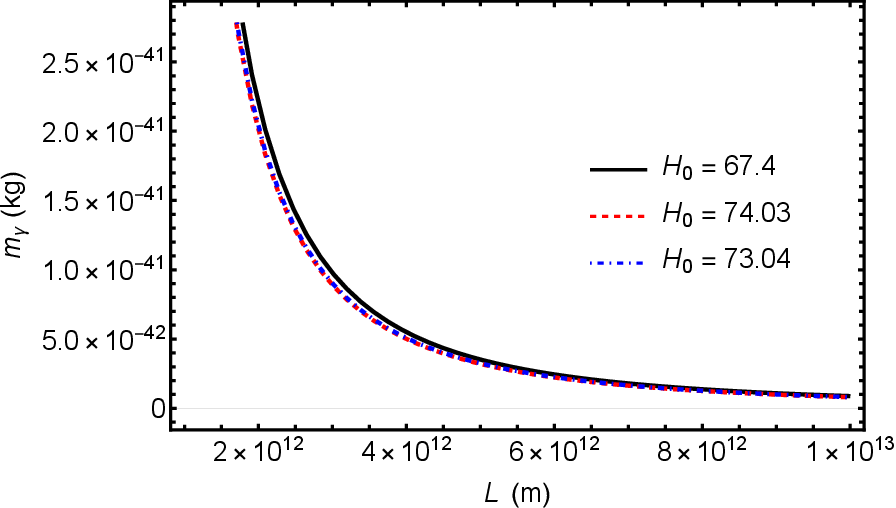}
\caption{\label{Fig1}\small{\emph{The illustration of the effective rest mass of photon in terms of $L$ for $H_{0}^{(\mathrm{PSM})}$ and $H_{0}^{(\mathrm{HST})}$ at $4.4\sigma$ and $5\sigma$ levels.}}}
\end{figure}

Figure \ref{Fig1} depicts the illustration of the effective rest mass of photon in terms of $L$ for $H_{0}^{(\mathrm{PSM})}$ and $H_{0}^{(\mathrm{HST})}$ at $4.4\sigma$, and at $5\sigma$ level based on Eq. \eqref{ermpnu}. From Fig. \ref{Fig1}, we see that increasing $L$ leads to reduce the effective rest mass of photon as seen from Table \ref{Table2}.

\subsection{Using $L_{\mathrm{H}}$ as the Hubble length to address the Hubble tension}

Now we aim to using $L_{\mathrm{H}}=\frac{c}{H_{0}}$ as the Hubble length to check whether this value for $L$ can reproduce the effective rest mass of photon and address the Hubble tension through the arithmetic mean change formula. To achieve this, by setting $L_{\mathrm{H}}=\frac{c}{H_{0}}$, we reevaluate Eqs. \eqref{amcmg}-\eqref{ermpnu} computed by using $H_{0}^{(\mathrm{PSM})}=67.4\,\,\mathrm{km\cdot s^{-1}\cdot Mpc^{-1}}$ \cite{Planck:2018vyg} with $H_{0}^{(\mathrm{HST})}=74.03\,\,\mathrm{km\cdot s^{-1}\cdot Mpc^{-1}}$ \cite{Riess:2019cxk} denoting the Hubble tension at $4.4\sigma$ level and also with $H_{0}^{(\mathrm{HST})}=73.04\,\,\mathrm{km\cdot s^{-1}\cdot Mpc^{-1}}$ \cite{Riess:2022jrx} describing the tension at $5\sigma$ level. The outcomes are shown in Table \ref{Table3}. From Table \ref{Table3}, we see that the deduced photon rest mass associated with $L_{\mathrm{H}}=\frac{c}{H_{0}}$ for various values of $H_{0}$ is very small (interestingly, with the same order of the photon rest mass achieved in the Capozziello et al. proposal \cite{Capozziello:2020nyq}) than the upper limits on the photon rest mass to the order of $\sim 10^{-54}\,\mathrm{kg}$ \cite{ParticleDataGroup:2022pth,Retino:2016gga} or $\sim 10^{-50}\,\mathrm{kg}$ \cite{Williams:1971ms}. Therefore, the Hubble length $L_{\mathrm{H}}$ cannot lead to predicted rest mass of photon. Moreover, we see from Table \ref{Table3} that increasing the value of $H_{0}$ leads to decreasing $L_{\mathrm{H}}$ and amplifying the effective rest mass of photon. Hence, the largest resulted value of the photon rest mass is related to the value of $H_{0}^{(\mathrm{HST})}=74.03\,\,\mathrm{km\cdot s^{-1}\cdot Mpc^{-1}}$ \cite{Riess:2019cxk} for the Hubble tension at $4.4\sigma$ level. On the other hand, the arithmetic mean change values demonstrate that although the reported rest mass of photon is not reproduced, the Hubble tension at both $4.4\sigma$ and $5\sigma$ can be addressed considering $L_{\mathrm{H}}$ as the EUP length scale in our setup due to the presence of large distances (IR) quantum gravity effects. From Tables \ref{Table2} and \ref{Table3}, we discover that apparently the value of $L$ for reproducing the rest mass of photon should be larger than $\sim 10^{13}\,\mathrm{m}$ and smaller than $\sim 10^{26}\,\mathrm{m}$.
\begin{table}[htb]
  \centering
  \caption{\label{Table3}\emph{The resulted photon rest mass and the arithmetic mean change associated with $L_{\mathrm{H}}=\frac{c}{H_{0}}$ as the Hubble length related to various values of $H_{0}$ in the EUP setup.}}
  \def\arraystretch{1.55}
  \begin{tabular}{l||l|l|l|l}
  \hline\hline
  $H_{0}\,\mathrm{(km\cdot s^{-1}\cdot Mpc^{-1})}$ & $L_{\mathrm{H}}=\frac{c}{H_{0}}\,\mathrm{(m)}$ & $m_{\gamma}\,\mathrm{(kg)}$ & $\frac{|\Delta m_{\gamma}|}{\bar{m}_{\gamma}}$ & $\frac{|\Delta H_{0}|}{\bar{H}_{0}}$ \\
  \hline
  $H_{0}^{(\mathrm{PSM})}=67.4$ \cite{Planck:2018vyg} & $L_{\mathrm{H}}^{(\mathrm{PSM})}\approx 1.373\times 10^{26}$ & $m_{\gamma}^{(\mathrm{PSM})}\approx 6.074\times 10^{-69}$ &  &  \\
  $H_{0}^{(\mathrm{HST})}=74.03\,(\mathrm{at}\,4.4\sigma)$ \cite{Riess:2019cxk} & $L_{\mathrm{H}}^{(\mathrm{HST})}\approx 1.250\times 10^{26}$ & $m_{\gamma}^{(\mathrm{HST})}\approx 6.671\times 10^{-69}$ & $\approx 0.09$ & $\approx 0.09$ \\
  $H_{0}^{(\mathrm{HST})}=73.04\,(\mathrm{at}\,5\sigma)$ \cite{Riess:2022jrx} & $L_{\mathrm{H}}^{(\mathrm{HST})}\approx 1.267\times 10^{26}$ & $m_{\gamma}^{(\mathrm{HST})}\approx 6.581\times 10^{-69}$ & $\approx 0.08$ & $\approx 0.08$ \\
  \hline\hline
\end{tabular}
\end{table}

\subsection{Constraining $\beta$ and $L$ through the effective rest mass of photon}\label{subsecc}

Here by considering Eq. \eqref{ermp}, we plan to utilize the aforementioned upper bounds on the effective rest mass of photon, i.e., $m_{\mathrm{\gamma}}<10^{-54}\,\mathrm{kg}$ \cite{ParticleDataGroup:2022pth,Retino:2016gga} and $m_{\mathrm{\gamma}}<1.6\times 10^{-50}\,\mathrm{kg}$ \cite{Williams:1971ms}, to constrain the EUP parameter $\beta=\frac{\beta_{0}}{L^{2}}$ and then, put some realistic limits on $L$ by assuming $\beta_{0}=1$. Therefore, we can insert values of numerical parameters in Eq. \eqref{ermp} considering $\beta=\frac{\beta_{0}}{L^{2}}$ to the form
\begin{equation}\label{ermpnuumm}
m_{\gamma}=6.433\times 10^{-52}H_{0}+\frac{1.922\times 10^{-34}\beta}{H_{0}}\,,
\end{equation}
which is in terms of $H_{0}$ and $\beta$.

For each of the aforementioned upper bounds on effective rest mass of photon, we find a limit on $\beta$ for both $H_{0}^{(\mathrm{PSM})}=67.4\,\,\mathrm{km\cdot s^{-1}\cdot Mpc^{-1}}$ \cite{Planck:2018vyg} with $H_{0}^{(\mathrm{HST})}=74.03\,\,\mathrm{km\cdot s^{-1}\cdot Mpc^{-1}}$ \cite{Riess:2019cxk} depicting the Hubble tension at $4.4\sigma$ level and with $H_{0}^{(\mathrm{HST})}=73.04\,\,\mathrm{km\cdot s^{-1}\cdot Mpc^{-1}}$ \cite{Riess:2022jrx} explaining the tension at $5\sigma$ level, respectively. Finally, we collect the outcomes in Table \ref{Table4}. From Table \ref{Table4}, we can discover that for $L\sim 10^{16}\,\mathrm{m}$, the upper limit $1.6\times 10^{-50}\,\mathrm{kg}$ on the photon rest mass \cite{Williams:1971ms} can be reproduced, while for $L\sim 10^{18}\,\mathrm{m}$, the upper bound $10^{-54}\,\mathrm{kg}$ on the photon rest mass \cite{ParticleDataGroup:2022pth,Retino:2016gga} may be generated. This observation proves that the presence of the EUP parameter $\beta$ can lead to reproducing the reported upper limits on the photon rest mass in this setup. Moreover, the deduced constraints on $L$ in Table \ref{Table4} demonstrate that the present setup can address the Hubble tension as an outcome of uncertainty limits in measuring the cosmological parameters.
\begin{table}[htb]
  \centering
  \caption{\label{Table4}\emph{The resulted bounds on $\beta$ and $L$ associated with various values of $m_{\gamma}$ and $H_{0}$ in the EUP setup.}}
  \def\arraystretch{1.55}
  \begin{tabular}{l|l||l||l|l}
  \hline\hline
  Test and Ref.   & $m_{\gamma}\,(\mathrm{kg})$ & $H_{0}\,\mathrm{(km\cdot s^{-1}\cdot Mpc^{-1})}$ & $\beta=\frac{\beta_{0}}{L^{2}}\,(\mathrm{m^{-2}})$ & $L\,(\mathrm{m})$ with $\beta_{0}=1$ \\
  \hline
  \multirow{3}{*}{Experimental tests \cite{Williams:1971ms}} & \multirow{3}{*}{$1.6\times 10^{-50}$} & $H_{0}^{(\mathrm{PSM})}=67.4$ \cite{Planck:2018vyg} & $1.818\times 10^{-34}$ & $7.417\times 10^{16}$ \\
                  &                             & $H_{0}^{(\mathrm{HST})}=74.03\,(\mathrm{at}\,4.4\sigma)$ \cite{Riess:2019cxk} & $1.997\times 10^{-34}$ & $7.076\times 10^{16}$                 \\
                  &                             & $H_{0}^{(\mathrm{HST})}=73.04\,(\mathrm{at}\,5\sigma)$ \cite{Riess:2022jrx}  & $1.970\times 10^{-34}$ & $7.125\times 10^{16}$                 \\
  \hline
  \multirow{3}{*}{Solar system data \cite{ParticleDataGroup:2022pth,Retino:2016gga}}      & \multirow{3}{*}{$10^{-54}$}       & $H_{0}^{(\mathrm{PSM})}=67.4$ \cite{Planck:2018vyg} & $1.136\times 10^{-38}$ & $9.382\times 10^{18}$ \\
                  &                             & $H_{0}^{(\mathrm{HST})}=74.03\,(\mathrm{at}\,4.4\sigma)$ \cite{Riess:2019cxk} & $1.248\times 10^{-38}$ & $8.951\times 10^{18}$                 \\
                  &                             & $H_{0}^{(\mathrm{HST})}=73.04\,(\mathrm{at}\,5\sigma)$ \cite{Riess:2022jrx}  & $1.231\times 10^{-38}$ & $9.013\times 10^{18}$                 \\
  \hline\hline
  \end{tabular}
\end{table}

\begin{figure}[H]
\centering
\includegraphics[width=0.7\textwidth]{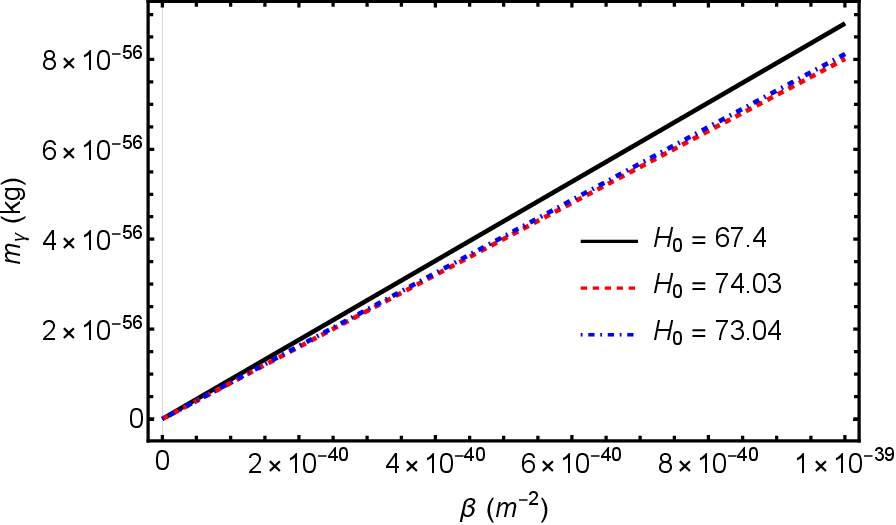}
\caption{\label{Fig2}\small{\emph{The graph of the effective rest mass of photon in terms of $\beta$ for $H_{0}^{(\mathrm{PSM})}$ and $H_{0}^{(\mathrm{HST})}$ at $4.4\sigma$, and at $5\sigma$ level.}}}
\end{figure}
Figure \ref{Fig2} illustrates the behavior of the effective rest mass of photon in terms of $\beta$ for $H_{0}^{(\mathrm{PSM})}$, $H_{0}^{(\mathrm{HST})}$ at $4.4\sigma$, and at $5\sigma$ level based on Eq. \eqref{ermpnuumm}. From Fig. \ref{Fig2}, we see that amplifying $\beta$ results in increasing the effective rest mass of photon.

\section{Concluding Remarks}\label{sec:4}

In this paper, we examined the possibility of addressing the Hubble tension in the framework of EUP (extended Uncertainty Principle) in which the late-time effects of the QG are considered. We followed the idea introduced by Capozziello et al. \cite{Capozziello:2020nyq} where in light of the HUP (Heisenberg Uncertainty Principle), the Hubble tension is explained as the consequence of uncertainties in measurements of cosmological parameters at cosmological scales. In this sense, we first found the EUP-modified reduced Compton wavelength of the low energy photons, which is in terms of $L$ as the EUP fundamental length denoting the trace of the QG effects at large distances scale. Apparently, this EUP-modified reduced Compton wavelength is shorter than the ordinary reduced Compton wavelength or even its GUP-modified version implemented by Trivedi \cite{Trivedi:2022bis}. Then, we equated the EUP-modified reduced Compton wavelength of the low energy photons to the luminosity distance truncated at the second order through the cosmographic approach. This, consequently, results in the expression for the effective rest mass of photon in terms of $H_{0}$, $z$, $q_{0}$, and interestingly the parameter $L$. This expression is the key tool deduced in our work to address the Hubble tension. We set $z=1$ where the kinematic interpretation of the Hubble-Lemaître law and the dynamical explanation derived from the Friedmann equations are equivalent.

Implementing the expression of the effective rest mass of photon, we planned to achieve at three major aims. First, we inserted the previously reported lower bounds on $L$ in the expression of the effective rest mass of photon for $H_{0}^{(\mathrm{PSM})}=67.4\,\,\mathrm{km\cdot s^{-1}\cdot Mpc^{-1}}$ \cite{Planck:2018vyg} with $H_{0}^{(\mathrm{HST})}=74.03\,\,\mathrm{km\cdot s^{-1}\cdot Mpc^{-1}}$ \cite{Riess:2019cxk} demonstrating the Hubble tension at $4.4\sigma$ level and also with $H_{0}^{(\mathrm{HST})}=73.04\,\,\mathrm{km\cdot s^{-1}\cdot Mpc^{-1}}$ \cite{Riess:2022jrx} describing the tension at $5\sigma$ level. Consequently, the rest mass of photon was found lower than the reported upper limit on the photon rest mass. However, we have found that the arithmetic mean changes of the photon rest mass and the Hubble-Lemaître parameter are approximately the same for $H_{0}^{(\mathrm{PSM})}$ and $H_{0}^{(\mathrm{HST})}$ such that at $4.4\sigma$ level, they both gained $\approx 0.09$ while at $5\sigma$ level, they both found as $\approx 0.08$. This fact demonstrated that in this EUP setup, the Hubble tension may be the effect of the uncertainties in measurements due to the QG effects in large-scale and there is no need to invoke new physics. Precisely speaking, the discrepancy between the photon rest mass associated with $H_{0}^{(\mathrm{PSM})}$ and $H_{0}^{(\mathrm{HST})}$ at $4.4\sigma$ and $5\sigma$ levels apparently is the cause of the Hubble tension. As the second purpose, we again repeated the first aim, now by utilizing the Hubble length as the value of the EUP parameter length $L$. As a result, the photon rest mass was deduced very much lower than the reported upper limit on the photon rest mass. However, through evaluating arithmetic mean changes of the photon rest mass and the Hubble-Lemaître parameter, we again found that the Hubble tension is due to the EUP limit as the QG large-scale effects in measuring cosmological parameters. As the third purpose, we utilized the reported upper bounds on the photon rest mass to constrain the EUP parameter length $L$ corresponding with each value of $H_{0}$. We found $L\sim 10^{16}\,\mathrm{m}$ associated with $1.6\times 10^{-50}\,\mathrm{kg}$ \cite{Williams:1971ms} and $L\sim 10^{18}\,\mathrm{m}$ related to $10^{-54}\,\mathrm{kg}$ \cite{ParticleDataGroup:2022pth,Retino:2016gga}.

According to the outcomes of the present study, we can conclude that the Hubble tension is a consequence of large-scale EUP effects, which influence the measurements of cosmological parameters. Moreover, it seems that regularizing the late-time epochs (or the large-scale structure; the IR limit) of the universe by the EUP setup leads to address the Hubble tension without having to invoke new physics, which apparently means that the Hubble tension itself perhaps is caused due to the lack of an appropriate theoretical explanation about the large-scale structure of the universe. We demonstrated that such an appropriate theoretical explanation for the large-scale structure of the universe can be the EUP setup indeed, which requires to be considered observationally in future observational studies. Additionally, we discovered that the QG effects at large distances scale described within the EUP setup apparently play the most important role in explaining cosmological and astrophysical phenomena. Finally, two major advantages of our setup compared to previous works are as follows. First, the EUP-modified reduced Compton wavelength is smaller than ordinary \cite{Capozziello:2020nyq} or GUP-revised one \cite{Trivedi:2022bis}, as mentioned-above, which able us to go beyond the HUP (or GUP) limit in addressing the tension as stated in the Capozziello et al. proposal \cite{Capozziello:2020nyq}. Second, rather considering early-time effects, e.g., GUP and compactified higher dimensions \cite{Trivedi:2022bis}, we took into consideration the late-time quantum gravity effects to address the Hubble tension since it is proved that just early-time new physics cannot resolve the Hubble tension \cite{Vagnozzi:2023nrq}.

\begin{acknowledgments}

The authors would like to thank the respectful referee for their insightful comments, which boosted the quality of the paper. The authors also appreciate Eleonora Di Valentino, Salvatore Capozziello, and Matteo Forconi for their insightful discussions and helpful advice in preparing the original draft of this paper.

\end{acknowledgments}

\end{document}